\nonstopmode
\documentclass{ECOC-author}

\usepackage{tikz}
\usepackage{pgfplots}
\usetikzlibrary{colorbrewer}
\usetikzlibrary{arrows}
\usetikzlibrary{shapes,decorations}
\usepackage{caption}
\usepackage{amsmath,amssymb}
\usepackage{mathtools, cuted}
\usepackage{graphicx}

\begin{document}

\title{Modulation Format Dependent, Closed-Form Formula for Estimating Nonlinear Interference in S+C+L Band Systems}

\author{Daniel Semrau\ad{1}\corr, Lidia Galdino\ad{1}, Eric Sillekens\ad{1}, Domani\c{c} Lavery\ad{1}, Robert I. Killey\ad{1}, Polina Bayvel\ad{1}}

\address{\add{1}{Optical Networks Group, Department of Electronic and Electrical Engineering, UCL (University College London), Torrington Place, London WC1E 7JE, United Kingdom}
\email{uceedfs@ucl.ac.uk}}

\begin{abstract}
A closed-form formula for the nonlinear interference estimation of arbitrary modulation formats in ultra-wideband transmission systems is presented. Enabled by the proposed approach, the formula is applied to the entire S+C+L band (20 THz) and validated by numerical simulations with experimentally measured fibre data.
\end{abstract}

\maketitle

\section{Introduction}
To support the ever increasing bandwidth demand, transmission over the entire S+C+L band is increasingly being considered \cite{exp0,exp1,exp2}. A major source of impairments over bandwidths beyond the conventional C-band is inter-channel stimulated Raman scattering (ISRS), which scales exponentially with optical bandwidth and imposes strong wavelength dependence on system performance. Formulas in closed-form, that yield performance estimates of all channels within seconds are essential for these bandwidth regimes enabling real-time network modeling, efficient link design and on-the-fly optimisation.
\par 
\ 
Extensions of the Gaussian Noise (GN) model to account for ISRS, termed the ISRS GN model, have been proposed in integral form \cite{OpEx, Roberts, Cantono, Cantono2, JLT, ECOC}. Numerically solving those integrals, typically taking a few minutes for a single transmission, quickly reaches unacceptable execution times, particularly for real-time optimisation and network modeling, where numerous configurations are analysed. To overcome this problem, modulation format independent closed-form approximations of the ISRS GN model have been proposed \cite{OpEx, cfJLT, Poggiolini}. Recently, we upgraded the closed-form formula to account for non-Gaussian signals, increasing the prediction accuracy for real-world modulation formats \cite{cfarxiv2}. However, fully analytical approaches are currently restricted to bandwidths of at most 15 THz, as there are currently no analytic solutions to the Raman equations for bandwidths beyond 15 THz \cite{ISRS}. The absence of such solutions prevents the derivation of closed-form formulas in this regime.
\par
\ 
In this paper, a modulation format dependent closed-form formula is presented together with a semi-analytical approach which allows the formula to be applied to optical bandwidths beyond 15 THz. The formula and the proposed approach are validated by split-step simulations over the entire S+C+L band (20~THz) using an experimentally measured Raman gain spectrum and attenuation profile. This work also represents the first validation of first-order perturbation models for optical bandwidths covering the entire S+C+L band.
\section{The ISRS GN model in closed-form}
\begin{figure*}[h]
   \centering
    \includegraphics[]{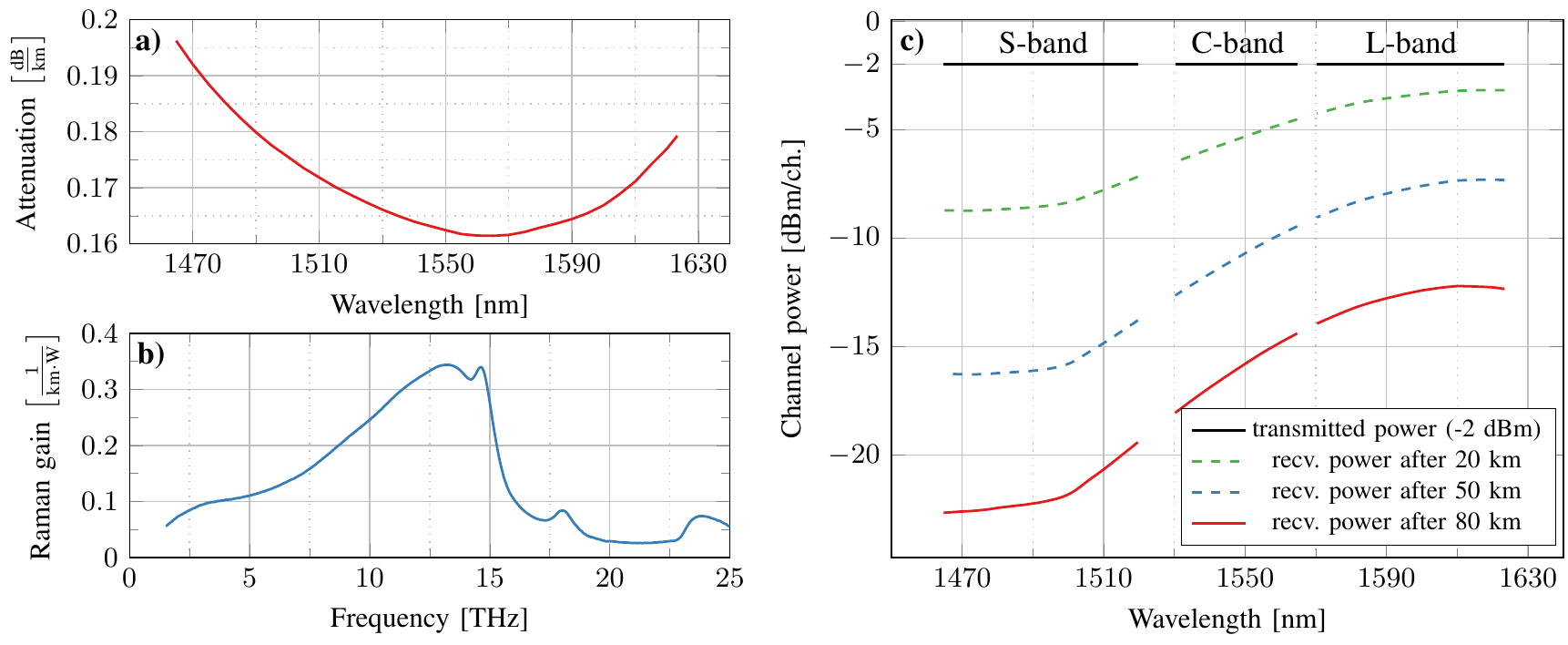}
\caption{\small \textcolor{black}{ Experimentally measured attenuation coefficient and Raman gain spectrum of a $\text{Corning}^{\text{\textcopyright}}$ $\text{SMF-28}^{\text{\textcopyright}}$ ULL fibre. The power after several propagation distances are shown in c).}}
\label{fig:fiber}
\end{figure*}
The total signal-to-noise ratio for an optical fibre communication system and a channel of interest $i$ is given by
\begin{equation}
\begin{split}
\text{SNR}_{\text{tot},i}^{-1} = \text{SNR}_{\text{NLI},i}^{-1} + \text{SNR}_{\text{ASE},i}^{-1}+ \text{SNR}_{\text{TRX},i}^{-1},
\label{eq:SNR}
\end{split}
\end{equation}
where $\text{SNR}_{\text{NLI},i}$, $\text{SNR}_{\text{ASE},i}$ and  $\text{SNR}_{\text{TRX},i}$ originate from fibre nonlinearity, amplifier noise and transceiver noise, respectively. This work focuses on the computation of $\text{SNR}_{\text{NLI},i}$.
\par
\ 
Recently, we presented a closed-form approximation of the ISRS GN model \cite{cfJLT} and extended the formula to account for arbitrary modulation formats \cite{cfarxiv2}. The formula is \eqref{eq:NLI} (see p. 2), where $T_i = \left(\alpha_i+\bar{\alpha}_i-P_{\text{tot}}C_{r,i}f_i\right)^2$, $A_i=\alpha_i+\bar{\alpha}_i$, $\phi=-4\pi^2\left[\beta_2+\pi\beta_3(f_i+f_k)\right]L$, $\phi_i=\frac{3}{2}\pi^2\left(\beta_2+2\pi\beta_3f_i\right)$, $\phi_{i,k}=-2\pi^2\left(f_k-f_i\right)\left[\beta_2+\pi\beta_3\left(f_i+f_k\right)\right]$, $\Phi$ is the excess kurtosis of the modulation format and $\tilde{n}=\left\{0 \ \text{for} \ n=1 \ \text{, } n \ \text{ for} \ n>1\right\}$. $P_i$ is the channel launch power with bandwidth $B_i$ centered at the relative frequency $f_i$, $P_{\text{tot}}$ is the total launch power and $n$ is the number of spans. The variables $\alpha_i$, $\bar{\alpha}_i$ and $C_{r,i}$ describe fibre loss and ISRS.
\par 
\ 
Eq. \eqref{eq:NLI} was derived assuming a linear regression of the Raman gain spectrum. This assumption yields an analytical solution of the Raman equations \cite{ISRS}; a key requirement to obtain a closed-form approximation for SNR in the presence of ISRS. For bandwidths beyond 15~THz, the Raman gain spectrum cannot be accurately approximated as a linear function of frequency and an analytical solution of the Raman equations is not available, rendering a fully analytical approach for SNR estimation impossible. To overcome this problem, we propose a semi-analytical approach, which yields vast reductions in computation time compared to split-step simulations.
\par 
\ 
To apply the proposed formula \eqref{eq:NLI} beyond 15~THz optical bandwidth, the loss and ISRS parameters $\alpha_i$, $\bar{\alpha_i}$ and $C_{r,i}$ are interpreted as channel dependent variables that are matched to the actual power profile of the respective channel $i$. For this purpose, the actual power profile $P_i\left(z\right)$ is obtained by numerically solving the Raman equations and matched to the power profile to first-order (with respect to ISRS)
\begin{equation}
\begin{split}
P^{\left(1\right)}_i\left(z\right)=\left(1+\tilde{T}_i\right)e^{-\alpha_i z} - \tilde{T}_ie^{-\left(\alpha_i+\bar{\alpha}_i\right) z},
\end{split}
\end{equation}
where $\tilde{T}_i  = -\frac{P_{\text{tot}}C_{r,i}}{\bar{\alpha}_i}f_i$. The matching of $\alpha_i$, $\bar{\alpha_i}$ and $C_{r,i}$ can be carried out using standard regression techniques.
\par 
\ 
This approach extends \eqref{eq:NLI} to bandwidths beyond 15 THz at the expense of numerically solving the Raman equations and additional regression operations which can be performed within a few seconds.
\par 
\
\begin{figure*}[b]
\vspace*{-0.5cm}
\begin{equation}
\begin{split}
\text{SNR}_{\text{NLI},i}^{-1}& \approx  \frac{4}{9}\frac{\pi \gamma^2P_i^2 n^{1+\epsilon} }{B^2_i\phi_{i}\bar{\alpha}_i\left(2\alpha_i+\bar{\alpha}_i\right)}\left[\frac{T_i-\alpha_i^2}{a_i}\text{asinh}\left(\frac{\phi_{i}B_i^2}{\pi a_i}\right)+\frac{A_i^2-T_i}{A_i}\text{asinh}\left(\frac{\phi_{i}B_i^2}{\pi A_i}\right)\right]\\
&+\frac{32}{27}\sum_{k=1,k\neq i}^{N_\mathrm{ch}} \frac{\gamma^2P_k^2}{B_k}\left\{\frac{n+\frac{5}{6}\Phi}{\phi_{i,k}\bar{\alpha}_k\left(2\alpha_k+\bar{\alpha}_k\right)}\right. \left[\frac{T_k-\alpha_k^2}{\alpha_k}\mathrm{atan}\left(\frac{\phi_{i,k}B_i}{\alpha_k}\right) +\frac{A_k^2-T_k}{A_k}\ \mathrm{atan}\left(\frac{\phi_{i,k}B_i}{A_k}\right)\right]\\
&\left.+\frac{5}{3} \frac{\Phi\pi \tilde{n} T_k }{\left|\phi\right| B_k^2\alpha_k^2 A_k^2}\left[\left(2\left|f_k-f_i\right|-B_k\right) \log\left(\frac{ 2\left|f_k-f_i\right| -B_k}{ 2\left|f_k-f_i\right| +B_k}\right) 2B_k\right]\right\}
\label{eq:NLI}
\end{split}
\end{equation}
\end{figure*}
\begin{figure*}[h]
   \centering
    \includegraphics[]{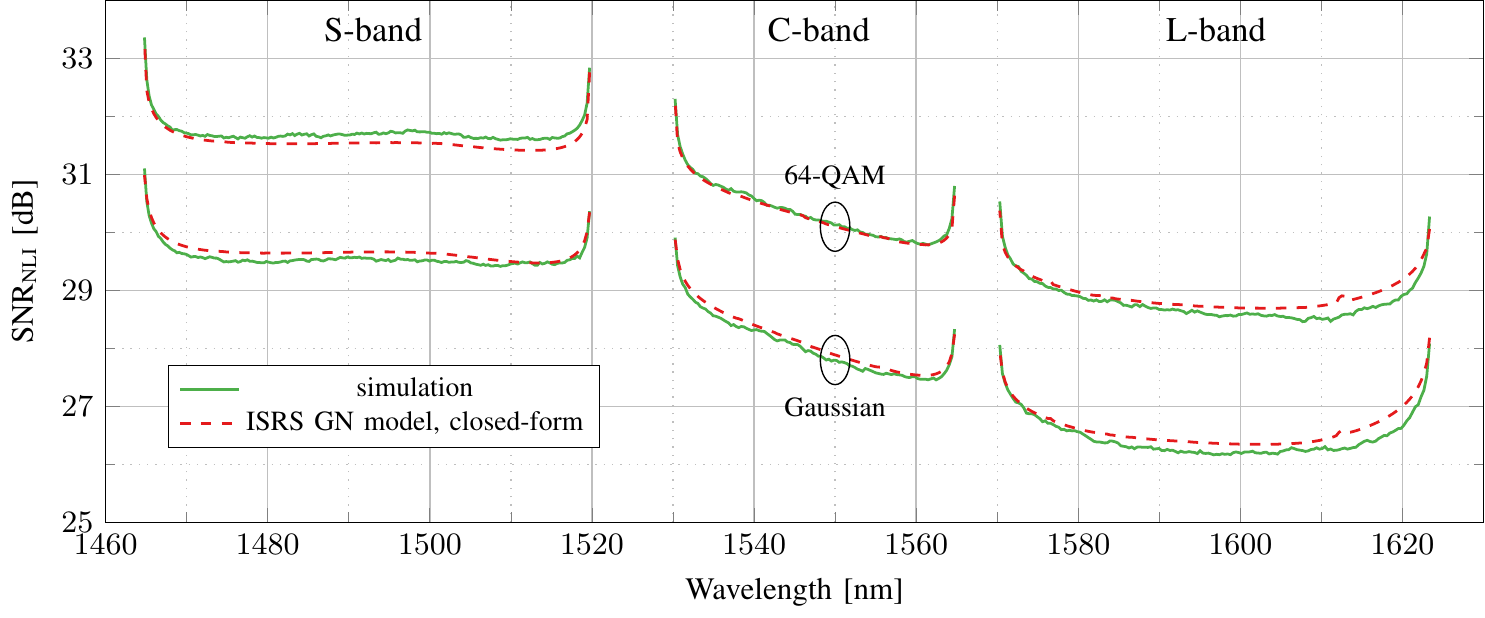}
\caption{\small Nonlinear performance after $\text{3}\times\text{80~km}$ transmission over the entire S+C+L band (20 THz, 158~nm) using a $\text{Corning}^{\text{\textcopyright}}$ $\text{SMF-28}^{\text{\textcopyright}}$ ULL with experimentally measured fibre data.}
\label{fig:NLI}
\end{figure*}
\vspace*{-0.5cm}
\section{Simulation setup}
To validate the proposed formula \eqref{eq:NLI}, a split-step simulation was performed for $\text{452}\times\text{40}$~GBd channels, occupying the entire S+C+L band (20~THz). This represents the first numerical validation of perturbation models such as \cite{Roberts, JLT, cfJLT, Cantono, cfarxiv2, Poggiolini} over optical bandwidths of 20~THz. A $\text{Corning}^{\text{\textcopyright}}$ $\text{SMF-28}^{\text{\textcopyright}}$ ULL fibre was considered with experimentally measured fibre data for attenuation coefficient, shown in Fig. \ref{fig:fiber}a), and the Raman gain spectrum, shown in Fig. \ref{fig:fiber}b). Dispersion and nonlinearity parameters are $D=18\frac{\text{ps}}{\text{nm}\cdot \text{km}}$, $S=0.067\frac{\text{ps}}{\text{nm}^2\cdot\text{km}}$ ($\beta_2=-22.6\frac{\text{ps}^2}{\text{km}}$ and $\beta_3=0.14\frac{\text{ps}^3}{\text{km}}$ at 1540~nm) and $\gamma=1.2\frac{\text{1}}{\text{W}\cdot\text{km}}$. 
\par 
\
The aim was to estimate the accuracy of the modulation format independent as well as the modulation format dependent NLI contribution of \eqref{eq:NLI}. For this reason, Gaussian as well as 64-QAM symbols were transmitted. A uniform channel launch power of -2~dBm was considered. A gap of 10~nm was assumed between S and C band and of 5~nm between C-and L-band.
\par 
\
A sequence length of $2^{16}$ symbols were found to be sufficient for accurate simulations of up to 3 fibre spans. However, to further increase the simulation accuracy, four independent data realisations were averaged. The step size was logarithmically distributed and ISRS was implemented as a frequency dependent loss at every step. Each data realisation took 4.2 days on a single state-of-the-art GPU.
\par 
\
The power after several propagation distances within a span is shown in Fig. 1c), where the signal power undergoes a complex interaction between the intrinsic fibre loss and ISRS. The sharp decrease of the Raman gain spectrum at frequency separations beyond 15 THz (approximately 120~nm) in combination with the non-uniform attenuation profile offset each other to give a relatively flat received power for low wavelengths in the S-band and high wavelengths in the L-band. Wavelengths in the C-band experience a linear tilt. This is a result of the coupling to all wavelengths in S- and L-band and the very low variation of the attenuation coefficient (maximum $0.04$~dB/km) within the C-band.
\section{Transmission results}
The nonlinear SNR as a function of wavelength after $\text{3}\times\text{80~km}$ is shown in Fig. \ref{fig:NLI}. As a sanity check, results obtained with the ISRS GN model in integral form \cite{JLT} are shown which exhibits a negligible error ($<$0.1~dB) compared to split-step simulations using Gaussian symbols. A Runge-Kutta method was used to numerically solve the Raman equations and a standard least mean squares algorithm was performed to minimise $\left|P_i\left(z\right)-P_i^{\left(1\right)}\left(z\right)\right|$ to obtain $\alpha_i$, $\bar{\alpha}_i$ and $C_{r,i}$.
\par 
\
Both the simulations and the models yield a relatively flat SNR across the wavelengths in S-and L-band. This is a consequence of the balancing effects between power tilt (see Fig. 1c)) and dispersion slope $S$ ($\beta_3$). The power tilt shown in Fig. 1c) results in increasing nonlinear interference towards longer wavelengths, while the dispersion slope weakens the nonlinear interference for longer wavelengths. Both effects result in a flat nonlinear SNR in the S-and L-bands for the given fibre parameters and the chosen launch power. The $\text{SNR}_\text{NLI}$ is tilted by 1.2~dB for wavelengths within the C-band as the power tilt occurring during propagation outweighs the effects of the dispersion slope. The nonlinear performance of channels within the S-band is about 3.3~dB higher as compared to channels in the L-band due to the large power depletion, originating from ISRS and wavelength dependent attenuation.
\par 
\ 
The proposed closed-form has remarkable accuracy with an average mismatch of 0.1~dB and a maximum mismatch of 0.3~dB in the case of Gaussian symbols. The formula correctly predicts the impact of non-Gaussian modulation formats such as 64-QAM with an average and maximum mismatch of 0.1 and 0.3 dB, respectively. This validates that \eqref{eq:NLI} can be applied in regimes beyond bandwidths of 15 THz with the approach described in this paper. The total simulation time to obtain the results in Fig. \ref{fig:NLI} was 33.6 days on a single state-of-the-art GPU. The total execution time using the semi-analytical closed-form approach was only a few seconds, where the majority of the time was required to numerically solve the Raman equations. This emphasises the major speed advantage of performance estimation approaches in closed-form.
\par 
\
The accuracy of \eqref{eq:NLI} is expected to be maintained for different launch power distributions, modulation formats, fibre types and longer transmission distances, making the formula particularly useful for metro and long-haul transmission distances, where ultra-wideband simulations become unmanageable.
\par 
\
The analysis shows that the proposed closed-form approximation can be applied to S+C+L band transmission systems using QAM formats and optical bandwidths beyond 15 THz. The formula can be universally applied to account for arbitrary launch power distributions, span numbers, modulation formats and ultra-wideband effects such as wavelength dependent attenuation, dispersion and inter-channel stimulated Raman scattering.
\section{Conclusion}
A closed-form formula was presented which accurately predicts the nonlinear performance of arbitrary modulation formats in ultra-wideband transmission. In combination with the proposed semi-analytical approach, the formula can be applied to optical bandwidths beyond 15 THz, where fully analytical solutions do not currently exist. Validations were carried out by numerical simulations over the entire S+C+L band, covering 20 THz optical bandwidth and using experimentally measured fibre data. The new method enables rapid performance estimation in S+C+L band transmission systems, allowing a more intelligent access to fibre capacity by correctly modelling optical fibre transmission properties.
\section{Acknowledgements}
Support for this work is from UK EPSRC under DTG PhD studentship to D.~Semrau, a RAEng research fellowship to D. Lavery and L. Galdino and EPSRC TRANSNET Programme Grant. The authors thank S. Makovejs from Corning for providing the experimental fibre data.

\end{document}